\documentclass[prd,twocolumn,graphicx,print]{revtex4-1}
\usepackage{graphicx}
\usepackage{dcolumn}
\usepackage{bm}
\usepackage{lipsum}
\usepackage{epsfig}
\usepackage{natbib}
\usepackage{amsmath}
\usepackage{mathrsfs}
\usepackage{amssymb}
 \usepackage{mathrsfs}
\usepackage{times}
\usepackage{psfrag}
\usepackage[english]{babel}
\usepackage[normal]{subfigure}
\usepackage{multirow}
\usepackage{color}
\begin{document}

\title{Geometrization of Scalar and Spinor Electrodynamics via Bohmian Quantum Gravity}

\author{Sijo K. Joseph}
\affiliation{Quantum Gravity Research, Topanga, CA 90290, USA.}

\begin{abstract}
Quantum theory is formulated as a probabilistic theory on a flat Minkowski space-time,  while general theory of relativity is formulated on a curved manifold as a geometric theory. Bohmian Quantum Gravity approach indicates that one need to convert a probabilistic theory to a geometric form to merge it with general theory
of relativity. We explore the differential geometric formulation of Scalar Electrodynamics and Spinor Electrodynamics and its coupling to the gravitational field equation. Using Feynman-Gell-Mann equation (second order Dirac equation), the fermionic matter field is nicely incorporated into general theory of relativity with the help of scalar-vector-tensor theory of gravity. Gravitationally coupled spin-field equations and generalized Feynman-Gell-Mann equation are derived from the action proposed here in the article. It is also shown that spin-electromagnetic interaction and spin-spin interaction can curve the space-time.
\end{abstract}

\date{\today}

\pacs{67}

\maketitle 

\section{Introduction}
There are many generalization to Einstein's theory of gravity~\cite{BeyondEinsteinGravity,WeylReview2017,Bergmann1968,fR_HamFormlation,fofT1,*fofT2,Bekenstein2011_TeVeS,
*BekensteinPRD_TeVeS,Moffat_STVG,*MOG_Moffat,Shojai_Article}, even though Scalar-Tensor Theory deserve a special consideration. Using extended version of Einstein's theory, namely the Scalar -Tensor theory of Gravity~\cite{Shojai2008},
it is found that the quantum theory can be incorporated into gravity in a geometric manner.
Scalar field contribution in the Scalar-Tensor Theory arises purely from a quantum mechanical quantity called
the quantum potential~\cite{Shojai_Article,GabayJoseph1,GabayJoseph2,SKJoseph1}. 
Such a geometrical unification of quantum theory to general theory of relativity is achieved using deBroglie-Bohm version of quantum theory~\cite{Bohm1975,BohmI,BohmII}.
In all these attempts, quantum mechanical Klein-Gordon matter field couples to gravity in a nice geometric fashion. 
More promising direction is to incorporate all the classical fields in a geometric frame work. Wheeler had started 
such a geometrodynamics program but incorporating fermionic matter field in this framework stood as a difficult 
problem\cite{GeometrodynReview}. 

Recent research on Klein-Gordon equation shows that a quantum mechanical wave equation has a correspondence to the conformally flat metric based geometric formulation~\cite{Shojai_Article,GabayJoseph1,GabayJoseph2}.
We can therefore explore this correspondence for Klein-Gordon-Maxwell-Einstein system in the same way the Klein-Gordon-Einstein System is formulated~\cite{Shojai_Article,GabayJoseph1}. This can be extended to 
Feynman-Gell-Mann-Einstein System in a nice manner using the recent results~\cite{Asenjo_Mahajan2015,Pesci2005,Petroni1984}.
It is to be suspected that, correspond to every quantum mechanical field equation formulated on a flat space-time has an equivalent geometrical counter part. One can always think about the correspondence between  geometric formulation of the theory and the quantum mechanical complex-field formulation. According to recent research, we need a  different physical interpretation of quantum theory to merge it with general theory of relativity. Quantum physics should be considered as something which emerge from a fluctuating vacuum. From the fluid dynamics perspective, there should exist a microscopic dynamics to make possible the fluid dynamics equations, and these equations emerge as a continuum limit. If this is true, there should be a fluctuating background field which provides energy to the particles. In Bohmian quantum gravity approach, we are compelled to accept the real existence of such a fluctuating background and the quantum mechanical laws naturally arises from this. A. I. Pesci and co-authors had shown that the quantum potential can be derived from the classical kinetic equations both for particles with and without spin \cite{Pesci2005}. Thus the quantum mechanical laws are emerging from a more fundamental statistical rules. In our previous article\cite{GabayJoseph1}, we have shown that the Lagrange multiplier appears in the Bohmian Quantum Gravity theory plays the role of such a fluctuating vacuum field. It is also shown that the uncertainty principle can emerge from such a fluctuating vacuum field~\cite{GabayJoseph1}. It is to be noted that, these results have connections to stochastic mechanics approach to quantum mechanics as proposed by Nelson~\cite{NelsonBook}. In his approach, quantum particles are driven by a kind of Brownian motion resulting from quantum fluctuation. Physical origin of such a fluctuation is still mysterious. In Bohmian-Quantum Gravity, such a fluctuating vacuum field  appears from the unification of gravitation and quantum mechanics  using deBroglie-Bohm version of Quantum theory~\cite{Bohm1975,BohmI,BohmII,PHollandBook,SheldonReview}. Bohmian-Quantum Gravity approach provides a natural avenue to incorporate all the  known classical fields in a geometric frame work. 

In this article, first we geometrize scalar and spinor electrodynamics and then couple it with gravitational field equations using scalar-vector-tensor Theory. In this formalism spin is included in a nice geometric manner using usual general relativity without introducing frame fields.
\section{Geometrization of Scalar Electrodynamics}
Let us take the Lagrangian density of the scalar electro-dynamics which has a local $U (1)$ symmetry and it is
given by,
\begin{eqnarray}
\mathscr{L}&=& (\partial_{\mu}\Phi+\frac{ie}{\hbar} {A}_{\mu}\Phi )(\partial^{\mu}{\Phi}^*-\frac{ie}{\hbar} {A}^{\mu}\Phi^* ) -\frac{m^2}{\hbar^2} \Phi\Phi^* \\ \nonumber
 & & -\frac{1}{4} {F}^{\mu\nu} {F}_{\mu\nu} \label{SEDLagrangian}.
\end{eqnarray}
Adopting the metric signature $(+,-,-,-)$,
the equation of motion of the charged scalar field composed of spinless particles is given by,
\begin{eqnarray}
(\partial_{\mu}+\frac{ie}{\hbar} {A}_{\mu})(\partial^{\mu}+\frac{ie}{\hbar} {A}^{\mu})\Phi+\frac{m^2}{\hbar^2} \Phi=0.
\nonumber \\
\end{eqnarray}


 
Taking $\Phi=\sqrt{\rho}\,e^{i/\hbar\,S(x,t)}$ and separating real and imaginary part, the Klein-Gordon equation coupled to electro-magnetic field becomes,

\begin{eqnarray}
(\partial_{\mu}{S}+e A_{\mu})(\partial^{\mu}{S}+e A^{\mu})&=&m^2\left(1+\frac{\hbar^2}{m^2}\frac{\partial_{\mu}\partial^ {\mu}\sqrt{\rho}}{\sqrt{\rho}}\right) \\
\partial_{\mu}\Bigl({\rho}(\partial^{\mu}S+e{A}^{\mu})\Bigr)&=&0 \label{emconti}
\end{eqnarray}


These equations can be written in a familiar form which is similar to previous works \cite{Shojai_Article,GabayJoseph1,GabayJoseph2}, 
\begin{eqnarray}
(\partial_{\mu}{S}+e A_{\mu})(\partial^{\mu}{S}+e A^{\mu})&=m^2\Omega^2 \label{HJem}\\
\partial_{\mu}\Big({\rho}(\partial^{\mu}S+e{A}^{\mu})\Big)=0 \label{ContiEq}
\end{eqnarray}
Here $\Omega^2=\exp{\left(\frac{\hbar^2}{m^2}\frac{\partial_{\mu}\partial^ {\mu}\sqrt{\rho}}{\sqrt{\rho}}\right)}\approx 1+\frac{\hbar^2}{m^2}\frac{\partial_{\mu}\partial^ {\mu}\sqrt{\rho}}{\sqrt{\rho}}$, where $Q=\frac{\hbar^2}{m^2}\frac{\partial_{\mu}\partial^ {\mu}\sqrt{\rho}}{\sqrt{\rho}}$ is the quantum potential. Note that Eq.~\ref{HJem} and Eq.~\ref{ContiEq} can easily be generalized to curved background. These equations (Eq.~\ref{HJem} and Eq.~\ref{ContiEq}) together determine the Klein-Gordon Equation coupled
to electro-magnetism in terms of the density $\rho$ and four momentum $\partial^{\mu}{S}$ of field variable $\Phi$ given in
Eq.~\ref{SEDLagrangian}. It can be easily seen that Eq.~\ref{HJem} describes the Hamilton-Jacobi equation for the charged scalar field $\Phi$ with a quantum correction to the mass term.
Following our previous works~\cite{GabayJoseph1,GabayJoseph2,SKJoseph1}, we can geometrize scalar electrodynamics problem by finding a metric corresponds to the equations of motion (see Eq.~\ref{HJem}).

We need to assume the exponential form of the conformal factor to get a positive definite 
effective mass $M_{eff}=m^2\Omega^2$, thus $\Omega^2=\exp{(Q)}$ or in other words $Q=\ln\Omega^2$ should be assumed. 
Hence we propose the following action for the geometrization of scalar electrodynamics,
\begin{eqnarray}
& &A[g_{\mu\nu},{\Omega}, S, \rho, A_\mu,\lambda]=\frac{1}{2k}\int{d^4x\sqrt{-g}\left(R\Omega^2-6\nabla_{\mu}\Omega\nabla^{\mu}\Omega\right)}\nonumber\\
& &+\int{d^4x\sqrt{-g} \left(\frac{\rho}{m}\Omega^2 (\nabla_{\mu}{S}+e A_{\mu})(\nabla^{\mu}{S}+e A^{\mu})-m\rho\Omega^4\right)} \nonumber\\
& &-\frac{1}{4}\int{d^4x\sqrt{-g}\, F_{\mu\nu} F^{\mu\nu}}  \nonumber\\
& &+\int{d^4x\sqrt{-g}\lambda 
\left[\ln(\Omega^2)-\left(\frac{\hbar^2}{m^2}\frac{\nabla_{\mu}\nabla^{\mu}\sqrt{\rho}}{\sqrt{\rho}}\right)\right]},
\end{eqnarray}
where the electro-magnetic field tensor is given by $F_{\mu\nu} = \nabla_\mu A_\nu - \nabla_\nu A_\mu$.
Minimizing this action with respect to different field variables will result into the following field equations.
Scalar curvature equation is given by,

\begin{eqnarray}
& & R\Omega+6 \Box\Omega +2\kappa\frac{\lambda}{\Omega} \nonumber \\ 
& & +\frac{2\kappa}{m}\rho \Omega \Bigr((\nabla_{\mu}{S}+e A_{\mu}) (\nabla^{\mu}{S}+e A^{\mu})-2m^2\Omega^2\Bigl)=0.\nonumber \\ \label{TraceEq}
\end{eqnarray}
Maxwell's equation can be found by varying  the action with respect to $A_{\mu}$,
\begin{eqnarray}
\nabla_{\mu}F^{\mu\nu}-\frac{1}{m}\rho\Omega^2(\nabla^{\nu}S+e A^{\nu}) =0. \label{MaxwellEq}  
\end{eqnarray}
Continuity equation is obtained from the variation of action with respect to $S$. This is just the imaginary part of the
quantum gravity corrected Klein-Gordon-Maxwell equation.
\begin{eqnarray}
\nabla_{\mu}\Bigl(\rho\Omega^2(\nabla^{\mu}S+e A^{\mu})\Bigr)=0 \label{GenEMContiEq}
\end{eqnarray}
The real part of the quantum gravity corrected Klein-Gordon-Maxwell equation is obtained by varying the action with respect to quantum mechanical density $\rho$,
\begin{eqnarray}
\Bigl((\nabla_{\mu}{S}+e A_{\mu}) (\nabla^{\mu}{S}+e A^{\mu})- m^2\Omega^2\Bigr) \Omega^2 \sqrt{\rho} \nonumber \\
+\frac{\hbar^2}{2m}[\Box({\frac{\lambda}{\sqrt{\rho}})}-\lambda\frac{\Box\sqrt{\rho}}{\rho}] =0. \label{EqEMDiracMotion}
\end{eqnarray}
Generalized Einstein's gravitational field equation for the matter-field without spin, which is coupled to electro-magnetism is obtained by varying the action with respect to the background metric tensor $g_{\mu\nu}$.
\begin{eqnarray}
& & \mathcal{G}_{\mu \nu} -\frac{[g_{\mu \nu}\Box- \nabla_{\mu}\nabla_{\nu}]\Omega^2}{\Omega^2}-6 \frac{\nabla_{\mu}\Omega \nabla_{\nu}\Omega}{\Omega^2} 
+3 g_{\mu \nu} \frac{\nabla_{\sigma}\Omega \nabla^{\sigma}\Omega}{\Omega^2} \nonumber \\ 
&& +\frac{2\kappa}{m}\rho (\nabla_{\mu}{S}+e A_{\mu})(\nabla_{\nu}{S}+e A_{\nu}) \nonumber \\ 
&& -\frac{\kappa}{m} \rho\,g_{\mu\nu} (\nabla_{\sigma}{S}+e A_{\sigma})(\nabla^{\sigma}{S}+e A^{\sigma})  
+ \kappa m \rho \Omega^2 g_{\mu\nu} \nonumber\\
&& -{\kappa}\left(F_{\mu\sigma}F_{\nu}^{\sigma}-\frac{1}{4}g_{\mu\nu}F_{\alpha\beta}F^{\alpha\beta}\right) \nonumber \\
&& +\frac{\kappa\hbar^2}{m^2\Omega^2}[\nabla_{\mu}\sqrt{\rho}\nabla_{\nu}(\frac{\lambda}{\sqrt{\rho}})
 +\nabla_{\nu}\sqrt{\rho}\nabla_{\mu}(\frac{\lambda}{\sqrt{\rho}})] \nonumber\\
&& -\frac{\kappa\hbar^2}{m^2\Omega^2}g_{\mu\nu}\nabla_{\sigma}(\lambda\frac{\nabla^{\sigma}{\sqrt{\rho}}}{\sqrt{\rho}})=0 \label{ScalarEinstein}
\end{eqnarray}

Constraint equation is given by,
\begin{eqnarray}
\Omega^2=\exp{\left(\frac{\hbar^2}{m^2}\frac{\nabla_{\mu}\nabla^ {\mu}\sqrt{\rho}}{\sqrt{\rho}}\right)}.
\end{eqnarray}

Equating the trace of the Eq.~\ref{ScalarEinstein} and Eq.~\ref{TraceEq}, $\lambda$-equation can also be found to be,

\begin{eqnarray}
\lambda=\frac{\hbar^2}{m^2}\frac{1}{(1-Q)}\nabla_{\mu}\Bigl(\lambda\frac{\nabla^{\mu}\sqrt{\rho}}{\sqrt{\rho}} \Bigr). \label{ExpLambdaEq}
\end{eqnarray}
In the linear order case ($\Omega^2=1+Q$), Eq.~\ref{ExpLambdaEq} becomes,
\begin{eqnarray}
\lambda=\frac{\hbar^2}{m^2}\nabla_{\mu}\Bigl(\lambda\frac{\nabla^{\mu}\sqrt{\rho}}{\sqrt{\rho}} \Bigr) \label{LinLambdaEq}.
\end{eqnarray}
Equation \ref{LinLambdaEq} is obtained by ignoring powers of $Q$ (keeping only $\hbar^2$ term) in the $\lambda$ expression given in Eq.~\ref{ExpLambdaEq}. This $\lambda$ field equation is already been interpreted as the vacuum field equation 
~\cite{GabayJoseph1,GabayJoseph2,SKJoseph1}.

\section{Geometrization of Spinor Electrodynamics}
It is a well known fact that, incorporating Dirac-matter field in general relativity is a difficult task. This can be easily achieved in this article using the Feynman-Gell-Mann equation and it is given by,
\begin{eqnarray}
(\partial^{\mu}+\frac{ie}{\hbar} {A}^{\mu})(\partial_{\mu}+\frac{ie}{\hbar} {A}_{\mu})\psi+\frac{e}{2\hbar}\sigma_{\mu\nu}F^{\mu\nu}\psi+\frac{m^2}{\hbar^2} \psi=0. \label{FeynGell}
\nonumber \\
\end{eqnarray}

It was N. C. Petroni, Ph. Gueret and J.-P.Vigier who had explored the hydrodynamical analysis
of the Feynman-Gell-Mann equation~\cite{Petroni1984}. These authors had derived the Feynman-Gell-Mann equation 
using Lagrangian formalism and the quantum potential is also determined. 
In Ref.~\cite{Pesci2005}, authors had derived the second-order version of the Dirac Equation using classical 
statistical arguments only. They have found that the Fourier transform of the one-particle distribution function of the classic relativistic Boltzmann equation with respect to the momentum variable can be mapped either onto 
the Klein-Gordon or the second order Dirac equation (Feynman-Gell-Mann equation). 
Recently in Ref.~\cite{Asenjo_Mahajan2015}, F. A Asenjo and S. M. Mahajan have constructed a fully relativistic quantum vortex dynamics of the hydrodynamical version of the Feynman-Gell-Mann equation. According Ref.~\cite{Pesci2005,Asenjo_Mahajan2015}, one can separate Eq.~\ref{FeynGell} to real and imaginary parts taking,
\begin{eqnarray}
\psi=\sqrt{{\psi}^{\dagger}\psi}\,e^{\frac{i}{\hbar}S}\,\begin{pmatrix} \cos{(\theta/2)}{e}^{i\eta/2} \\ 
i\sin{(\theta/2){e}^{-i\eta/2}} \end{pmatrix} \label{HydroPsi}.
\end{eqnarray}
Here the extra $\theta$ and $\eta$ variables are incorporated to take into account the spinor nature of the two-component wave-function $\psi$. The quantum mechanical density is defined as $\rho={\psi}^{\dagger}\psi$. The four component Dirac-Spinor is given by, $\Psi=\begin{pmatrix}\, \psi \\ -\psi \end{pmatrix}$. 
This  decomposition conserves the spinor degrees of freedom in terms of two spin-fields $\theta$ and $\eta$.
These spin-fields $\theta$ and $\eta$ constitute a parametric representation of the unimodular spin vector with spin components $\Sigma_{1} = \sin\theta \sin\eta $, $\Sigma_{2} = \sin\theta \cos\eta$, and $\Sigma_{3} = \cos\theta$, therefore, $\eta = \arctan(\Sigma_{1}/\Sigma_{2})$. Define $q(\zeta) = {\hbar^2}/{4}-{\zeta}^{2}$ 
and  $\zeta = \hbar/2\, \Sigma_{3}=\hbar/2\cos\theta$, then defining the most general four-momentum $\mathcal{P}_{\mu}$ by incorporating the spin variables $\zeta$ and $\eta$, 
\begin{eqnarray}
\mathcal{P}_{\mu}=(\partial_{\mu}{S}+e A_{\mu}+\zeta \partial_{\mu}\eta).
\end{eqnarray}
Substituting Eq.~\ref{HydroPsi} and separating real and imaginary part, the Feynman-Gell-Mann equation becomes,

\begin{eqnarray}
\mathcal{P}_{\mu}\mathcal{P}^{\mu} & = & m^2\Bigl(1+\frac{\hbar^2}{m^2}\frac{\Box{\sqrt{\rho}}}{\sqrt{\rho}} 
 +\frac{e\hbar}{2m^2}M_{\alpha\beta}\,F^{\alpha\beta} \nonumber \\
& & -\frac{\hbar^2}{4m^2}\partial_{\mu}M^{\alpha\beta}\partial^{\mu}{M^{*}}\!\!_{\alpha\beta}\Bigr)\label{SpinEqMotion}
\end{eqnarray}

\begin{eqnarray}
\partial_{\mu}\Bigl(\rho\mathcal{P}^{\mu}\Bigr)=0 \label{SpinContiEq}
\end{eqnarray}
Considering Eq.~\ref{SpinEqMotion} and following the arguments of deBroglie ~\cite{deBroglieBook},
F. Shojai ~\cite{Shojai_Article} and others \cite{GabayJoseph1,GabayJoseph2,SKJoseph1}, one can find a differential manifold with a metric $\tilde{g}_{\mu\nu}$ where the particle with spin moves
on this manifold freely. Particle follows the geodesic on the differential manifold defined by the metric
$\tilde{g}_{\mu\nu}$. Hence quantum theory can be geometrized and can be coupled with gravity in a nice manner 
once the quantum potential $Q$ or the conformal factor $\Omega^2$ is known. All the previous works along this direction didn't consider quantum mechanical particles with spin. In this manuscript, we exploit the hydrodynamic picture of the Dirac equation via the Feynman-Gell-mann equation in order to couple it with the gravitational field. Aforementioned Eq.~\ref{SpinEqMotion} and Eq.~\ref{SpinContiEq} can be re-arranged in a much more familiar form, making similar to previous works \cite{Shojai_Article,GabayJoseph1,GabayJoseph2,SKJoseph1}, 
\begin{eqnarray}
\mathcal{P}_{\mu}\mathcal{P}^{\mu}=m^2\Omega_{D}^2 \label{DiracHJeq} \\
\partial_{\mu}\Bigl(\rho\mathcal{P}^{\mu}\Bigr)=0. \label{DiracContiEq}
\end{eqnarray}
Here $\Omega_{D}^2=1+Q_{D}$, where $Q_{D}=Q_{KG}+Q_{spin-em}+Q_{spin-spin}=\frac{\hbar^2}{m^2}\frac{\Box\sqrt{\rho}}{\sqrt{\rho}}+\frac{e\hbar}{2m^2}M_{\alpha\beta}\,F^{\alpha\beta}-\frac{\hbar^2}{4m^2}\partial_{\mu}M^{\alpha\beta}\partial^{\mu}{M^{*}}\!\!_{\alpha\beta}$, where $Q_{D}$ is termed as the Dirac-Quantum Potential. Here $\Omega_{D}^2$ can be termed as the Dirac-Shojai conformal factor. This Dirac-Shojai conformal factor defines a curved space-time $\tilde{g}_{\mu\nu}=\Omega_{D}^2\eta_{\mu\nu}$ which is conformally flat. The space-time curvature generated by the quantum mechanical density $\rho$ is already known~\cite{Shojai2007,Shojai2008,Shojai_Article,GabayJoseph1,GabayJoseph2,SKJoseph1}. Here two important additional effects appears, (a) Spin-tensor $M_{\alpha\beta}$  couples with electro-magnetic tensor $F^{\alpha\beta}$ and yields a curvature contribution, (b) space-time variation of the spin-tensor $M_{\alpha\beta}$ interact with its conjugate and generates an opposite curvature effect. Physical implications of such a curvature effect from the spin-spin interaction term $-\frac{\hbar^2}{4m^2}\partial_{\mu}M^{\alpha\beta}\partial^{\mu}{M^{*}}\!\!_{\alpha\beta}$ will be worthwhile for our future explorations. In addition to the equation of motion (Eq.~\ref{DiracHJeq}) and continuity equation (Eq.~\ref{DiracContiEq}), two spin-field equations can also be obtained, see Ref.~\cite{Pesci2005} and Ref~.\cite{Asenjo_Mahajan2015} for details.
Equation~\ref{DiracHJeq}, Eq.~\ref{DiracContiEq} and two-spin-field equations together determine Feynman-Gell-Mann Equation  in terms of the spinor density $\rho$, four momentum $\partial^{\mu}{S}$ and spin-field variables $\zeta$ 
and $\eta$. Going back to Eq.~\ref{DiracHJeq}, it can be easily seen that Eq.~\ref{DiracHJeq} describes the Hamilton-Jacobi equation for the spinor field with a quantum correction to the mass term including a spin contribution.

As discussed earlier, following our previous work, we can geometrize the spinor electrodynamics problem by finding 
a metric $\tilde{g}_{\mu\nu}=\Omega_{D}^2\eta_{\mu\nu}$ corresponds to the equations of motion (see Eq.~\ref{DiracHJeq}) which is determined by the Dirac-Shojai conformal factor $\Omega_{D}^2$. 
Dirac-Shojai conformal factor defines a conformally flat metric in the absence of gravity (when the background metric is $\eta_{\mu\nu}$).
Hence we take the following action with a curved background metric $g_{\mu\nu}$ (see Eq.~\ref{FGAction}) in order to geometrize quantum mechanical matter field with spin and couple it with gravitational field equations,
\begin{widetext}
\begin{eqnarray}
& & A[g_{\mu\nu},{\Omega_{D}}, S, \rho,\eta,\zeta, A_\mu,\lambda]= 
\frac{1}{2k}\int{d^4x\sqrt{-g}\Bigl(R\Omega_{D}^2-6\nabla_{\mu}\Omega_{D}\nabla^{\mu}\Omega_{D}\Bigr)}  \nonumber \\
& & +\int d^4x\sqrt{-g} \Bigl(\frac{\rho}{m}\Omega_{D}^2 (\nabla_{\mu}{S}+e A_{\mu}+\zeta \nabla_{\mu}\eta)
(\nabla^{\mu}{S}+e A^{\mu}+\zeta \nabla^{\mu}\eta)-m\rho\Omega_{D}^4\Bigr)  \nonumber \\
& & -\frac{1}{4}\int{d^4x\sqrt{-g}\, F_{\mu\nu} F^{\mu\nu}}  \nonumber \\
& & +\int{d^4x\sqrt{-g}\lambda \left[\ln(\Omega_{D}^2)-\left(\frac{\hbar^2}{m^2}\frac{\Box{\sqrt{\rho}}}{\sqrt{\rho}}
    +\frac{e\hbar}{2m^2}{M}_{\alpha\beta}\,F^{\alpha\beta}-\frac{\hbar^2}{4m^2}\Bigl(\frac{\nabla^{\mu}\zeta \nabla_{\mu}\zeta}{q(\zeta)}+\frac{4}{\hbar^2}q(\zeta){\nabla^{\mu}\eta \nabla_{\mu}\eta}\Bigr) \right)\right]},\nonumber \\ \label{FGAction}
\end{eqnarray}
\end{widetext}

where the electro-magnetic field tensor is given by $F_{\mu\nu} = \nabla_\mu A_\nu - \nabla_\nu A_\mu$.
Taking the identity by Pesci et.al~\cite{Pesci2005} and generalizing that to a curved manifold, the following identity holds
\begin{equation}
\nabla_{\mu} M^{\alpha\beta}\nabla^{\mu}{M^{*}}\!\!_{\alpha\beta}=\frac{\nabla_{\mu}\zeta \nabla^{\mu}\zeta}{q(\zeta)}+\frac{4}{\hbar^2}q(\zeta){\nabla_{\mu}\eta \nabla^{\mu}\eta} 
\end{equation}
where $q(\zeta)=\frac{\hbar^2}{4}-\zeta^2$. Here the spin-tensor $M^{\alpha\beta}$ is defined as $M_{\alpha\beta}=\frac{\Psi^{\dagger}\sigma_{\alpha\beta}\Psi}{\Psi^{\dagger}\Psi}$ also $\sigma_{\alpha\beta}=\frac{i}{2}(\gamma_{\alpha}\gamma_{\beta}-\gamma_{\beta}\gamma_{\alpha})$.
Minimizing the action given in Eq.~\ref{FGAction} with respect to different field variables will result into the following field equations.

Scalar curvature equation with spin contribution is obtained by varying the action (Eq.~\ref{FGAction}) with respect to the conformal factor $\Omega_{D}$,
\begin{eqnarray}
& & R\,\Omega_{D}+6\,\Box\Omega_{D} +2\kappa\,\frac{\lambda}{\Omega_{D}}\nonumber \\ 
& & +\frac{2\kappa}{m} \rho\, \Omega_{D} \Bigr((\nabla_{\mu}{S}+e A_{\mu}+\zeta \nabla_{\mu}\eta) 
(\nabla^{\mu}{S}+e A^{\mu}+\zeta \nabla^{\mu}\eta) \nonumber \\
& & -2m^2\Omega_{D}^2\Bigl)=0\nonumber \\ \label{TraceEq}
\end{eqnarray}
Maxwell's equation can be found by varying  the action (Eq.~\ref{FGAction}) with respect to $A_{\mu}$,
\begin{eqnarray}
\nabla_{\mu}F^{\mu\nu}-\frac{1}{m}\rho\Omega_{D}^2(\nabla^{\nu}S+e A^{\nu}+\zeta \nabla^{\nu}\eta) \nonumber \\
+\frac{e\hbar}{m^2}\nabla_{\mu}(\lambda\,M^{\mu\nu}) =0 \label{MaxwellEq}  
\end{eqnarray}
Continuity equation is obtained from the variation of action (Eq.~\ref{FGAction}) with respect to $S$. 
This is just the imaginary part of the
quantum gravity corrected Feynman-Gel-Mann-Maxwell equation.
\begin{eqnarray}
\nabla_{\mu}\Bigl(\rho\Omega_{D}^2(\nabla^{\mu}S+e A^{\mu}+\zeta \nabla^{\mu}\eta)\Bigr)=0 \label{GenContiDiracEq}
\end{eqnarray}
The real part of the quantum-gravity corrected Feynman-Gel-Mann-Maxwell equation is obtained by varying the action (Eq.~\ref{FGAction}) with respect to quantum mechanical density $\rho$,
\begin{eqnarray}
& & \Bigl[(\nabla_{\mu}{S}+e A_{\mu}+\zeta \nabla_{\mu}\eta) (\nabla^{\mu}{S}+e A^{\mu}+\zeta \nabla^{\mu}\eta)- m^2\Omega_{D}^2\Bigr] \nonumber \\
& & +\frac{\hbar^2}{2m \Omega_{D}^2 \sqrt{\rho}}[\Box({\frac{\lambda}{\sqrt{\rho}})}-\lambda\frac{\Box\sqrt{\rho}}{\rho}] =0 \label{EqEMDiracMotion}
\end{eqnarray}
Generalized Einstein's field equation for the fermionic matter-field, which is coupled to electro-magnetism is obtained by varying the action with respect to the background metric tensor $g_{\mu\nu}$.
\begin{eqnarray}
\mathcal{G}_{\mu \nu} -\frac{[g_{\mu \nu}\Box- \nabla_{\mu}\nabla_{\nu}]\Omega_{D}^2}{\Omega_{D}^2}-6 \frac{\nabla_{\mu}\Omega \nabla_{\nu}\Omega}{\Omega_{D}^2} 
+3 g_{\mu \nu} \frac{\nabla_{\sigma}\Omega \nabla^{\sigma}\Omega}{\Omega_{D}^2} \nonumber \\ 
+\frac{2\kappa}{m}\rho (\nabla_{\mu}{S}+e A_{\mu}+\zeta \nabla_{\mu}\eta)(\nabla_{\nu}{S}+e A_{\nu}+\zeta \nabla_{\nu}\eta) \nonumber \\ 
-\frac{\kappa}{m} \rho\,g_{\mu\nu} (\nabla_{\sigma}{S}+e A_{\sigma}+\zeta \nabla_{\sigma}\eta)(\nabla^{\sigma}{S}+e A^{\sigma}+\zeta \nabla^{\sigma}\eta)  \nonumber\\
+ \kappa m \rho \Omega_{D}^2 g_{\mu\nu} \nonumber\\
-\frac{\kappa}{\Omega_{D}^2}\left(F_{\mu\sigma}F_{\nu}^{\sigma}-\frac{1}{4}g_{\mu\nu}F_{\alpha\beta}F^{\alpha\beta}\right) \nonumber \\
+\frac{\kappa\hbar^2}{m^2\Omega_{D}^2}[\nabla_{\mu}\sqrt{\rho}\nabla_{\nu}(\frac{\lambda}{\sqrt{\rho}})
+\nabla_{\nu}\sqrt{\rho}\nabla_{\mu}(\frac{\lambda}{\sqrt{\rho}})] \nonumber\\
-\frac{\kappa\hbar^2}{m^2\Omega_{D}^2}g_{\mu\nu}\nabla_{\sigma}(\lambda\frac{\nabla^{\sigma}{\sqrt{\rho}}}{\sqrt{\rho}})\nonumber\\ +\frac{\kappa\hbar^2\lambda}{4m^2\Omega_{D}^2}(\nabla_{\mu}M^{\alpha\beta}\nabla_{\nu}{M^{*}{\!\!}}_{\alpha\beta}+\nabla_{\nu}M^{\alpha\beta}\nabla_{\mu}{M^{*}{\!\!}}_{\alpha\beta}) \nonumber \\
-\frac{\kappa\hbar^2\lambda}{4m^2\Omega_{D}^2} g_{\mu\nu}\nabla_{\sigma}M^{\alpha\beta}\nabla^{\sigma}{M^{*}{\!\!}}_{\alpha\beta} \nonumber \\ 
-\frac{\kappa\,e\,\hbar}{2m\Omega_{D}^2}\left(M_{\mu\sigma}F_{\nu}^{\sigma}+M_{\nu\sigma}F_{\mu}^{\sigma}-g_{\mu\nu}M_{\alpha\beta}F^{\alpha\beta}\right) =0, \nonumber \\
\end{eqnarray}
where $\mathcal{G}_{\mu \nu}$ is the well known Einstein tensor $\mathcal{G}_{\mu\nu} = R_{\mu\nu} - {1\over2} g_{\mu\nu}R$. The constraint equation is given by,
\begin{eqnarray}
\Omega_{D}^2&=&\exp{(Q_{D})},
\end{eqnarray}
where the quantum potential is given by,
\begin{eqnarray}
Q_{D}&=&\Biggr(\frac{\hbar^2}{m^2}\frac{\Box{\sqrt{\rho}}}{\sqrt{\rho}}
    +\frac{e\hbar}{2m^2}{M}_{\alpha\beta}\,F^{\alpha\beta} \nonumber \\
    & & -\frac{\hbar^2}{4m^2}\nabla_{\mu} M^{\alpha\beta}\nabla^{\mu}{{M^{*}}\!\!}_{\alpha\beta}\Biggl).
\end{eqnarray}

In addition to all these field equations, spin variables $\zeta$ and $\eta$ yields two spin-field equations which is coupled to gravity. Gravitationally coupled generalized spin-field equations are obtained by varying the action (see Eq.~\ref{FGAction}) with respect to $\eta$ and $\zeta$, and the following Eq.~\ref{spin_gr_eq1} and Eq.~\ref{spin_gr_eq2} are obtained,

\begin{eqnarray}
& & \Omega_{D}^{2} (\nabla_{\mu}{S}+e A_{\mu}+\zeta \nabla_{\mu}\eta)\nabla^{\mu}{\zeta} \nonumber \\
& & +\frac{1}{m\rho}\nabla_{\mu}\Bigl(\lambda\, q(\zeta)\,\nabla^{\mu}\eta\Bigr)+\frac{e\hbar}{4m}\frac{\lambda}{\rho}\frac{\partial{M}_{\alpha\beta}}{\partial\eta}\,F^{\alpha\beta}=0, \nonumber \\ \label{spin_gr_eq1}
\end{eqnarray}
\begin{eqnarray}
& & \Omega_{D}^2 (\nabla_{\mu}{S}+e A_{\mu} +\zeta \nabla_{\mu}\eta)\nabla^{\mu}{\eta}\nonumber \\
& & -\frac{\zeta\lambda}{m\rho} \nabla_{\mu}\eta \nabla^{\mu}{\eta} +\frac{\hbar^2}{4m}\frac{\zeta}{q(\zeta)^2} \frac{\lambda}{\rho}\,\nabla_{\mu}\zeta \,\nabla^{\mu}\zeta  \nonumber \\
& & -\frac{\hbar^2}{4m\rho}\nabla_{\mu}\Bigl( \frac{\lambda}{q(\zeta)}\,\nabla^{\mu}\zeta \Bigr)
-\frac{e\hbar}{4m} \frac{\lambda}{\rho} \frac{\partial{M}_{\alpha\beta}}{\partial\zeta}\,F^{\alpha\beta}=0. \label{spin_gr_eq2}
\end{eqnarray}
It can be easily seen that the spin-field equations coupled to gravity and vacuum (See Eq.~\ref{spin_gr_eq1} and Eq.~\ref{spin_gr_eq2}) and it can reproduce the equations given in Ref~\cite{Asenjo_Mahajan2015,Pesci2005} for a specific
choice of the Lagrange multiplier $\lambda=\lambda_{0}\rho$, where $\lambda_{0}$ is just a constant. Note that equations will be slightly different in coefficients since we use a different action. 
In our previous work~\cite{GabayJoseph2}, it is shown that the specific choice $\lambda=\lambda_{0}\rho$ can make vanish the vacuum coupling contributions in Klein-Gordon equation. In such a scenario, one can deal with the usual Klein-Gordon equation without any quantum-gravity corrections. The same situation is applicable in the case of Feynman-Gell-Mann equation.
Taking $\lambda=\rho$ in Eq.~\ref{spin_gr_eq1} and Eq.~\ref{spin_gr_eq2}, we get vacuum de-coupled spin-field equations,
\begin{eqnarray}
\mathcal{P}^{\mu}\nabla_{\mu}\zeta+\frac{1}{m\rho}\nabla_{\mu}\Bigl(\rho\, q(\zeta)\,\nabla^{\mu}\eta\Bigr)+\frac{e\hbar}{4m}\frac{\partial{M}_{\alpha\beta}}{\partial\eta}\,F^{\alpha\beta}=0, \nonumber \\ \label{free_spin_gr_eq1}
\end{eqnarray}
\begin{eqnarray}
& & \mathcal{P}^{\mu}\nabla^{\mu}{\eta}-\frac{\zeta}{m} \nabla_{\mu}\eta \nabla^{\mu}{\eta} +\frac{\hbar^2}{4m}\frac{\zeta}{q(\zeta)^2} \nabla_{\mu}\zeta \,\nabla^{\mu}\zeta  \nonumber \\
& & -\frac{\hbar^2}{4m\rho}\nabla_{\mu}\Bigl( \frac{\rho}{q(\zeta)}\,\nabla^{\mu}\zeta \Bigr) 
-\frac{e\hbar}{4m} \frac{\partial{M}_{\alpha\beta}}{\partial\zeta}\,F^{\alpha\beta}=0, \label{free_spin_gr_eq2}
\end{eqnarray}
and the equation of motion becomes,
\begin{eqnarray}
(\nabla_{\mu}{S}+e A_{\mu}+\zeta \nabla_{\mu}\eta) (\nabla^{\mu}{S}+e A^{\mu}+\zeta \nabla^{\mu}\eta)-m^2\Omega_{D}^2 =0. \nonumber \\  \label{FreeEqEMDiracMotion}
\end{eqnarray}
Ignoring quantum-gravity effect by taking $\Omega^2=1$, continuity equation becomes,
\begin{eqnarray}
\nabla_{\mu}\Bigl(\rho(\nabla^{\mu}S+e A^{\mu}+\zeta \nabla^{\mu}\eta)\Bigr)=0. \label{FreeContiDiracEq}
\end{eqnarray}
These four equations (Eq.~\ref{free_spin_gr_eq1}, Eq.~\ref{free_spin_gr_eq2}, Eq.~\ref{FreeEqEMDiracMotion} and Eq.~\ref{FreeContiDiracEq} ) can be mapped back to usual Feyman-Gell-Mann equation given in Eq.~\ref{FeynGell}.

An important contribution of energy momentum tensor arises due to the spin-electro-magnetic interaction and spin-spin interaction. This can be captured in an entirely new stress-energy tensor $T^{spin}_{\mu\nu}(\rho,\lambda,\zeta,\eta,A^{\mu})$
where new interesting Physics appears. The extra stress-energy tensor is given by,

\begin{eqnarray}
& & T^{spin}_{\mu\nu}(\rho,\lambda,\zeta,\eta,A^{\mu})= \nonumber \\
& & \frac{\kappa\hbar^2\lambda}{4m^2\Omega_{D}^2}(\nabla_{\mu}M^{\alpha\beta}\nabla_{\nu}{M^{*}{\!\!}}_{\alpha\beta}+\nabla_{\nu}M^{\alpha\beta}\nabla_{\mu}{M^{*}{\!\!}}_{\alpha\beta}) \nonumber \\
& & -\frac{\kappa\hbar^2\lambda}{4m^2\Omega_{D}^2} g_{\mu\nu}\nabla_{\sigma}M^{\alpha\beta}\nabla^{\sigma}{M^{*}{\!\!}}_{\alpha\beta} \nonumber \\
& & -\frac{\kappa\,e\,\hbar}{2m\Omega_{D}^2}\left(M_{\mu\sigma}F_{\nu}^{\sigma}+M_{\nu\sigma}F_{\mu}^{\sigma}-g_{\mu\nu}M_{\alpha\beta}F^{\alpha\beta}\right).
\end{eqnarray}
As a side remark, this spin-stress-energy tensor contains terms involving vacuum-spin-spin interaction and vacuum-spin-electromagnetic interaction term. Physical application of such interactions and its gravitational implication will be worthwhile to explore in the future work. Apart from that, one can include a topological term like $F_{\mu\nu} \tilde{F}^{\mu\nu}$ in the action given in Eq.~\ref{FGAction} , which will give the same generalized Einstein equation but vacuum field equations will be different and an electro-magnetic source term can appear in the $\lambda$ equation.

%

\section{Geometrization of Maxwell's Equations}

While minimizing the Lagrangian with respect to the vector potential $A_{\mu}$, we will get 
the following Maxwell's equation,
\begin{eqnarray}
\nabla_{\mu}F^{\mu\nu}-\frac{1}{m}\rho\Omega^2(\nabla^{\nu}S+e A^{\nu}+\zeta \nabla^{\mu}\eta) \nonumber\\
+\frac{e\hbar}{m^2}\nabla_{\mu}(\lambda\,M^{\mu\nu}) =0 \label{MaxwellEq2}.
\end{eqnarray}
Hence the Maxwell's equation written in terms of current density $J^{\nu}$ becomes,
\begin{eqnarray}
\nabla_{\mu} F^{\mu\nu} =\Omega^2 J^{\nu}-\frac{e\hbar}{m^2}\nabla_{\mu}(\lambda\,M^{\mu\nu}). \label{MaxwellEq}
\end{eqnarray}
where $J^{\nu}=\frac{\rho}{m}(\nabla^{\nu}S+e A^{\nu}+\zeta \nabla^{\mu}\eta)$. Here ${J_{\mathbf{vs}}}^{\nu}= -\frac{e\hbar}{m^2}\nabla_{\mu}(\lambda\,M^{\mu\nu})$ can be seen as an extra vacuum-spin current due to the coupling of spin-tensor 
$M^{\mu\nu}$ with the vacuum field $\lambda$, this can be seen as a small quantum-gravity corrections to the Maxwell's equation.

Due to the antisymmetric nature of $F^{\mu\nu}$ and $M^{\mu\nu}$, one can show that $\nabla_{\mu}\nabla_{\nu}F^{\mu\nu}=0$ and $\nabla_{\mu}\nabla_{\nu}M^{\mu\nu}=0$,
thus the current density equation becomes, 
\begin{eqnarray}
\nabla_{\mu}\Bigr(\Omega^2 J^{\mu}\Bigl)=0
\end{eqnarray}

Since $J^{\nu}=\frac{\rho}{m}(\nabla^{\nu}S+e A^{\nu}+\zeta \nabla^{\mu}\eta)$, it can be shown that, 
\begin{eqnarray}
\nabla_{\mu}\Bigl(\rho\Omega^2(\nabla^{\nu}S+e A^{\nu}+\zeta \nabla^{\mu}\eta)\Bigr)=0
\end{eqnarray}
This is the continuity equation of the particles with spin in the presence of the electro-magnetic field and gravitational coupling ($\Omega^2$ correction). It is already implied that the vacuum-spin current
${J_{\mathbf{vs}}}^{\nu}$ is separately conserved i.e. $\nabla_{\nu}{J_{\mathbf{vs}}}^{\nu}=0$. Note that, Eq.~\ref{GenContiDiracEq} is obtained by varying the action (given Eq.~\ref{FGAction}) with respect to the quantum mechanical density $\rho$.
In addition, the expression given here is more general than the quantum-mechanical continuity equation studied in previous works~\cite{Shojai_Article,GabayJoseph1,GabayJoseph2,GabayJoseph2}. Here the gravitational coupling with electro-magnetic field is achieved in an indirect manner via the conformal factor $\Omega^2$ in the current conservation equation (See Eq.~\ref{GenContiDiracEq}).
It is worthwhile to note that ${J_{s}}^{\mu}=\frac{1}{m}\rho\Omega^2 \zeta \nabla^{\mu}\eta$ gives the spin current contribution.

\section{Generalized Feynman-Gell-Mann Equation} 
Due to the presence of extra $\Omega_{D}^2$ in the generalized continuity equation (see Eq.~\ref{GenContiDiracEq}) as compared to Eq.~\ref{emconti} or Eq.~\ref{DiracContiEq}, we will always end up with a wave equation containing dissipation contribution as a conformal gravity correction. Similarly the equation of motion (Eq.~\ref{EqEMDiracMotion}) contains an extra vacuum-correction term $f(\lambda,\rho)=\frac{\hbar^2}{2m \Omega_{D}^2 \sqrt{\rho}}[\Box({\frac{\lambda}{\sqrt{\rho}})}-\lambda\frac{\Box\sqrt{\rho}}{\rho}]$. This term correction term $f(\lambda,\rho)$ acts like a forcing contribution arises to balance the dissipation contribution coming from the $\Omega_{D}^2$ factor appearing in the continuity equation (see  Eq.~\ref{GenContiDiracEq}). Thus, combining Eq.~\ref{GenContiDiracEq} and Eq.~\ref{EqEMDiracMotion} into a single complex equation, one can obtain a generalized Feynman-Gell-Mann equation in the wave-function picture with quantum-gravity correction and it is 
found to be, 
\begin{eqnarray}
&&\Bigr(\nabla_{\mu}+\frac{ie}{\hbar} {A}_{\mu}\Bigl)\Bigr(\nabla^{\mu}+\frac{ie}{\hbar} {A}^{\mu}\Bigl)\psi
+\frac{e}{2\hbar}\sigma_{\mu\nu}F^{\mu\nu}\psi\nonumber\\
&& +\frac{m^2}{\hbar^2}\psi+\frac{i}{\hbar}\Bigl(\frac{\nabla_{\mu}\Omega^2}{\Omega^2}{(\nabla^{\mu}S+eA^{\mu}+\zeta \nabla^{\mu}\eta)}\Bigr)\psi \nonumber \\
& & -\frac{1}{2m \Omega_{D}^2 \sqrt{\rho}}[\Box({\frac{\lambda}{\sqrt{\rho}})}-\lambda\frac{\Box\sqrt{\rho}}{\rho}]\psi =0. \label{QGRFGMeqn1} 
\end{eqnarray}
Applying the constraint condition $\Omega^2=e^{Q_{D}}$, and using the identity $\zeta \nabla^{\mu}\eta=-i\hbar{\psi}^{\dagger}\nabla^{\mu}\psi$ the equation simplifies to,
\begin{eqnarray}
& & \Bigr(\nabla_{\mu}+\frac{ie}{\hbar} {A}_{\mu}\Bigl)\Bigr(\nabla^{\mu}+\frac{ie}{\hbar} {A}^{\mu}\Bigl)\psi +\frac{e}{2\hbar}\sigma_{\mu\nu}F^{\mu\nu}\psi+\frac{m^2}{\hbar^2}\psi \nonumber \\ 
& & +\frac{i}{\hbar}\Bigl({\nabla_{\mu}Q_{D}}{(\nabla^{\mu}S+eA^{\mu})}\Bigr)\psi +(\nabla_{\mu}Q_{D}{\psi}^{\dagger}\nabla^{\mu}\psi)\psi \nonumber \\
& & -\frac{1}{2m \Omega_{D}^2 \sqrt{\rho}}[\Box({\frac{\lambda}{\sqrt{\rho}})}-\lambda\frac{\Box\sqrt{\rho}}{\rho}]\psi =0,   \label{QGRFGMeqn2} 
\end{eqnarray}
this is the quantum-gravity corrected Feynmann-Gell-Mann equation. It is to be noted that we are naturally lead to a nonlinear extension of quantum theory with almost similar results as in Ref.~\cite{Mahajan2015}.
Here the extra nonlinear term appears as a correction due to the vacuum energy and a dissipating contribution arises as the conformal gravity correction. The added advantage of finding the complex wave equation (see Eq.\ref{QGRFGMeqn2}) is that, we can easily see
how the quantum-gravity corrections appears in the usual quantum mechanical formalism. When the quantum mechanical matter density $\rho$ conforms with the fluctuating vacuum field $\lambda$, one reaches to an equilibrium situation which is determined by a linear differential equation. Quantum mechanical linear differential equations appears when the dissipation term in the theory balances with the forcing term. It is interesting to note that, $\lambda=\rho$ and $\Omega^2=1$, Eq.~\ref{QGRFGMeqn2} simplifies to,
\begin{eqnarray}
& & \Bigr(\nabla_{\mu}+\frac{ie}{\hbar} {A}_{\mu}\Bigl)\Bigr(\nabla^{\mu}+\frac{ie}{\hbar} {A}^{\mu}\Bigl)\psi +\frac{e}{2\hbar}\sigma_{\mu\nu}F^{\mu\nu}\psi+\frac{m^2}{\hbar^2}\psi =0, \nonumber \\   \label{QWFeqn1} 
\end{eqnarray}
which is the original Feynman-Gell-Mann equation.

\section{Conclusions}

In this geometrical formalism we have seen that scalar-electrodynamics can be geometrized using the scalar-vector-tensor theory, where the quantum mechanical conformal factor encodes the scalar matter field dynamics. 
Here we have also found the correspondence between the spinor-electrodynamics to its corresponding geometric counter part which is also a scalar-vector-tensor theory with two extra spin-field equations, where the vector field is just electromagnetism and geometrical scalar-field is the quantum mechanical matter-field. These results will shed light on the understanding of the General relativity and Quantum theory in a geometrical frame-work. Here we claim that classical field theories can be casted either in the form of fields on a flat manifold or it can be fully geometrized in terms of a curved manifold. Here the gravitational scalar field or the conformal factor $\Omega^2$ appears both from the Quantum Mechanical Klein-Gordon Field and the Feynman-Gell-Mann equation. In the case of spin, two extra spin-field equation coupled to gravity is present and an extra spin-energy-momentum tensor is also present. It is also shown that spin-electromagnetic interaction and spin-spin interaction can curve the space-time, since it appears in the conformal factor $\Omega_{D}^2$.  It is evident that, corresponds to every classical complex field equation formulated on a flat space-time, there is an equivalent curved geometry counter part.

%

\end{document}